# Robust CS reconstruction based on appropriate minimization norm

Maja Lakićević, Mitar Moračan, Nađa Đerković
Faculty of Electrical Engineering
Podgorica, Montenegro

*Abstract*-**Noise robust compressive sensing algorithm is considered. This algorithm allows an efficient signal reconstruction in the presence of different types of noise due to the possibility to change minimization norm. For instance, the commonly used $l_1$ and $l_2$ norms, provide good results in case of Laplace and Gaussian noise. However, when the signal is corrupted by Cauchy or Cubic Gaussian noise, these norms fail to provide accurate reconstruction. Therefore, in order to achieve accurate reconstruction, the application of $l_3$ minimization norm is analyzed. The efficiency of algorithm will be demonstrated on examples.**
*Keywords*-**Compressive sensing, signal reconstruction, minimization norms, sparse, non-iterative algorithm**

I. INTRODUCTION

The applications and algorithms, based on the signals sampled according to the Niquist sampling rate, may require significant hardware resources, which can further make them expensive and complex. A solution has been found in Compressive Sensing (CS), a method that allows the analysis and reconstruction of the signal, from the small set of random samples [1]-[4]. An important conditions for successful reconstruction are signal sparsity in one of the transformation domains (such as DFT, DCT, Wavelet domain etc.). Various algorithms have been proposed for signal reconstruction [5]-[9]. However, the situation becomes complicate in the presence of noise, and the algorithm should be modified to correspond to the nature of noise dealing with [10]-[14]. One possibility is to adapt the norm minimization problem to the noise nature [10],[11], or to apply the noise removing algorithms [13],[14] prior to the reconstruction.

The algorithm considered in this paper provides simple reconstruction solution using non-iterative approach, and especially offers a possibility to choose and change different minimization norms in order to obtain good reconstruction results [11]. Moreover, the approach in [11] established the relationship between robust statistics and the CS. The considered relationship between these two concepts is based on the initial robust formulations of the signal transforms and the property that incomplete set of samples causes random deviations of the DFT outside the signal frequencies. In addition, the sum of generalized deviations of the values at non-signal frequencies is higher than at the signal components positions. Therefore, we have to determine the threshold which will select signal components in order to provide good reconstruction. The main advantage of this algorithm is the possibility to use different of minimization norms, which is not the case in the most of the existing algorithms. For instance, in [11], it was shown that the signal reconstruction in the presence of impulsive and Gaussian noise, using $l_1$ and $l_2$ minimization norms, respectively. Here, we consider another types of noise, such as Cauchy and Cubic Gaussian noises. Moreover, we show that in this case $l_1$ and $l_2$ norms fail, but the accurate results can be achieved using the $l_3$ minimization norm.

The paper is organized as follows. The theoretical background is given in Section II, while Section III shows full mathematical fundaments of used algorithm. Results and possibilities of algorithm are presented in Section IV. Conclusion is presented in Section V.

II. THEORETICAL BACKGROUND

According to the compressive sensing theory signal *x* or its DFT vector *X*, can be reconstructed, with a high probability, from an incomplete set of measurements *y*, by solving a convex optimization problem [11]. This approach can be closely related to the robust transformation theory based on the modelling and minimization of certain error function. Consequently, let us observe a total error in the form:

$$E = \sum_{n=0}^{N-1} |e(n)|^L \qquad (1)$$

where error *e* can be defined as:

$$e(n) = x(n)\exp^{-\frac{j2\pi kn}{N}} - \mu \qquad (2)$$

and *X*(*k*), for *k* = 0,1,2,…,*N*, is the Fourier transform of the signal *x*(*n*). Minimizing total error by *X*(*k*):

$$\frac{\partial E}{\partial \mu} = 0 \big|_{\mu = X(k)} \quad (3)$$

we can obtain optimal FT for different type of noises. For example the standard definitions of the transform domain representations for *M* measurements is obtained as the solution of the above optimization problem for $|e|^2$:

$$X(k) = \frac{1}{N} \sum_{n=0}^{N-1} x(n) e^{-j2\pi kn/N} \quad (4)$$

Function $|e|^2$ is obtained from principle of estimation using maximum likelihood (ML). Based on ML estimation function, $|e(n)|^L$ can be determined as [14]:

$$|e(n)|^L = -\log(p_{noise}(e)) \quad (5)$$

where $p_{noise}$ is probability density function (pdf) of noise *noise(n)*. For noise with Gauss pdf function $p_{noise}(e) \sim \exp(-|e|^2)$, function $|e|^2$ represents ML estimator. However, for impulse noise like Laplace which pdf function have form $p_{noise}(e) \sim \exp(-|e|)$, function $|e|$ represents ML estimator. Transform domain representations in this case for M measurements, is obtained using $|e|$ and has form:

$$X(k) = median\{x(n) e^{-j2\pi kn/N}\} \quad \text{for } k=0,..,M. \quad (6)$$

In the case of compressive sampled signal *x(n)* the number of available samples *M* is much fewer than *N* so in sequel we will only consider *M* of *N* total sample. After calculating the error values for each available sample, based on function $|e(n)|^L$, we can calculate the sum of general deviations for each frequency [11]:

$$GD(k) = \frac{1}{M} \sum_{n=0}^{M-1} \left| \left( x(n) \exp^{-\frac{j2\pi kn}{N}} - \frac{1}{M} \sum_{n=0}^{M-1} x(n) \exp^{-\frac{j2\pi kn}{N}} \right) \right|^L \quad (7)$$

In the case when the function is $|e|^2$ GD is equal variance. From [11] GD is in this case also equal:

$$GD(k = k_j) = \frac{M(M-N)}{N-1} \sum_{i=1, i \neq j}^{K} A_i^L \quad (8)$$

$$GD(k \neq k_j) = \frac{M(M-N)}{N-1} \sum_{i=1}^{K} A_i^L \quad (9)$$

where $k=k_j$ is frequency of signal component and $A_i$ is amplitude of signal. Note that changing *L* we obtain different minimization norm. Norm $l_1$ is obtained with $(n) = |e|$, while norm $l_2$ is obtained with $e(n) = |e|^2$. In case of impulse noise is already proved that norm $l_1$ gives the best signal reconstruction because of form of its pdf function. The $l_3$−norm is obtained with $e(n) = |e|^3$. In sequel it will be tested how minimization of function $|e|^3$ influence on signal with different noises except already tested Gauss and Laplace.

### III. ALGORITHM

In this sequel, the algorithm for CS reconstruction of sparse signals will be described. The algorithm will be described in few steps [11]:
1. For each *k*=0,1,..,*N* calculate *X(k)* and GD(k) for $l_3$ norm which use $|e|^3$ function.
2. Determine the position of the all minimum values in order to remove the highest component $A_i$.

$$k_j = \arg\{GD(k) < \theta\}, \text{ for } k=1,...,N, \quad (10)$$

where *T* is threshold. *T* can be calculated considering $\max\{V(k)\}$ ,e.g., $\alpha \max\{V(k)\}$ where α represents a constant between 0.85 and 0.95. The appropriate value for α we obtained experimentally. Threshold can be also calculated considering $\alpha mean\{V(k)\}$ or $\alpha median\{V(k)\}$.

In CS matrix formed from DFT matrix, we keep only the rows that correspond to the extracted frequencies $k_{Oi}$ and columns coresponding to the available measurements $n_m$. In this way we obtain indeterminate system of equations $y = A_{CS}X$, which can be defined as:

$$X = (A_{CS}^* A_{CS})^{-1} A_{CS}^* y \quad (11)$$

The reconstructed amplitudes $A_i$ of coefficients in *X*, containing initial phases $\varphi_i$, are accurate for all $k_{Oi}$.

### IV. EXPERIMENTAL RESULTS

The sinusoidal signal, with length of 128 samples, is considered. It is sparse in DFT domain, so we use it for signal reconstruction, based on CS. The signal has three components and has the folowing form:

$$x(n) = A_1 e^{\left(\frac{j2\pi 16n}{N}\right)} + A_2 e^{\left(\frac{j2\pi 32n}{N}\right)} + A_3 e^{\left(\frac{j2\pi 64n}{N}\right)} \quad (12)$$

where $A_1$=4, $A_2$=3, $A_3$=2 are amplitudes of signal *x(n)*. Since these components behave in a similar manner, the intention is to estimate all components at once, without using the iterative procedure, from a small number of random signal measurements. The number of measurementsused for reconstruction and parameter α are constant (*M*=64 and α=0.89). Other parameters vary from one to another example so we will explain it later.

*Example 1:* In this case we considered the signal *x(n)* corrupted by Couchy noise:

$$noise = \frac{\sigma_1 \cdot randn(1,N) + j \cdot \sigma_2 \cdot randn(1,N)}{\sigma_1 \cdot randn(1,N) + j \cdot \sigma_2 \cdot randn(1,N)} \quad (13)$$

where $\sigma_1 = 1$ and $\sigma_2 = 1$. Reconstruction of this noisy signal is tested with different minimization norms using algoritm described previously.

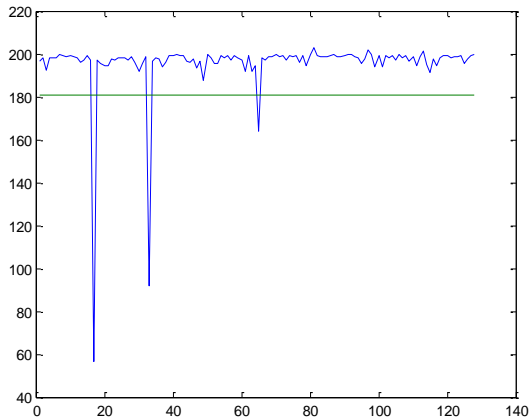

Fig. 1. GD with $|e|^3$ minimization function (blue line) and treshold $T = 0.89 * \max\{V(k)\}$ (green line)

We also obtained GD using $l_2$ and $l_1$ minimization norms. The algorithm gives different results for different minimization norms as we can see on Fig.2, Fig.3. By analyzing the result from Fig.2 and Fig.3 conclude that $l_3$ minimization norm provides best result and most credible reconstruction. If we run that same algoritm with same parameter 30 times $l_2$ will give better result comparing to norm $l_1$ but worse comparing to norm $l_3$.

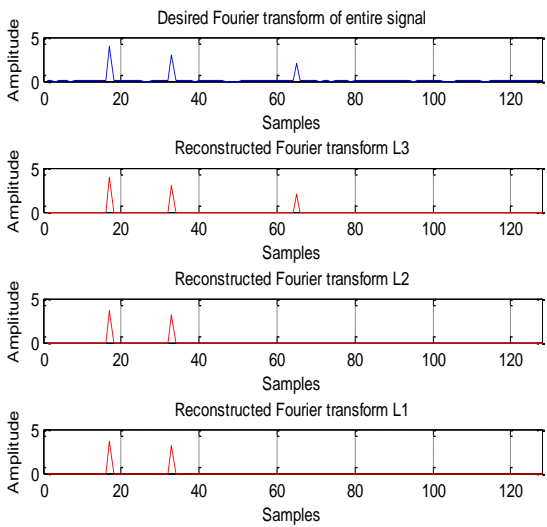

Fig. 2. Desired and reconstructed signal in FT domain using $l_3$, $l_2$ and $l_1$ minimization norms

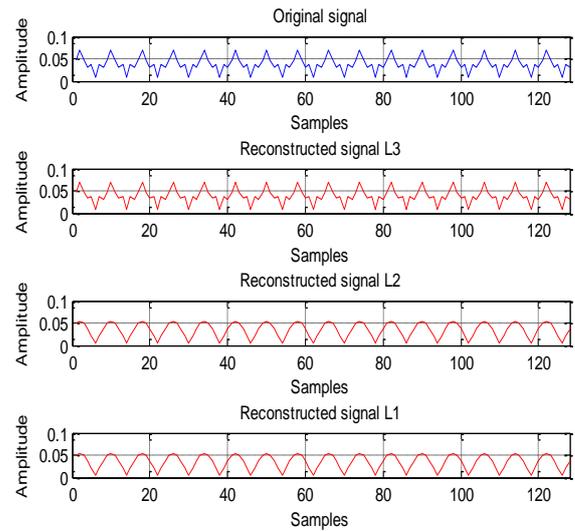

Fig. 3. Original and reconstructed signal in time domain using $l_3$, $l_2$ and $l_1$ minimization norms

*Example 2:* Here, we considered an example with the signal corrupted by this specific noise:

$$noise = (\sigma_1 \cdot randn(1,N))\wedge 3 + j \cdot (\sigma_2 \cdot randn(1,N))\wedge 3 \quad (14)$$

where $\sigma_1 = 1$ and $\sigma_2 = 1$.

Reconstruction of noisy signal is tested with different minimization norms. For this specific noise results are presented on Fig.4 and Fig.5.

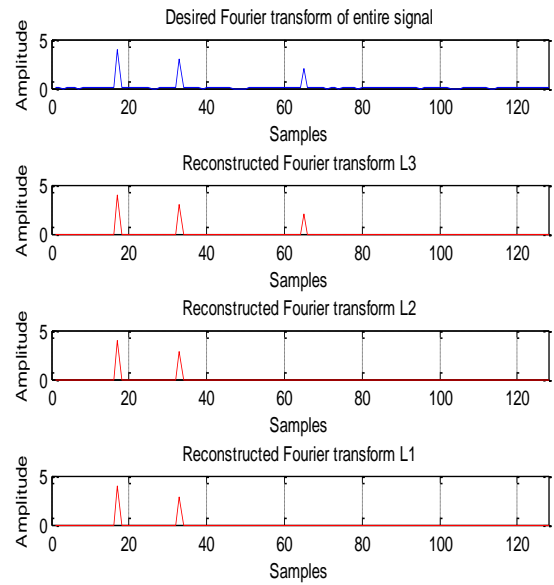

Fig. 4. Desired and reconstructed signal in FT domain using $l_3$, $l_2$ and $l_1$ minimization norms

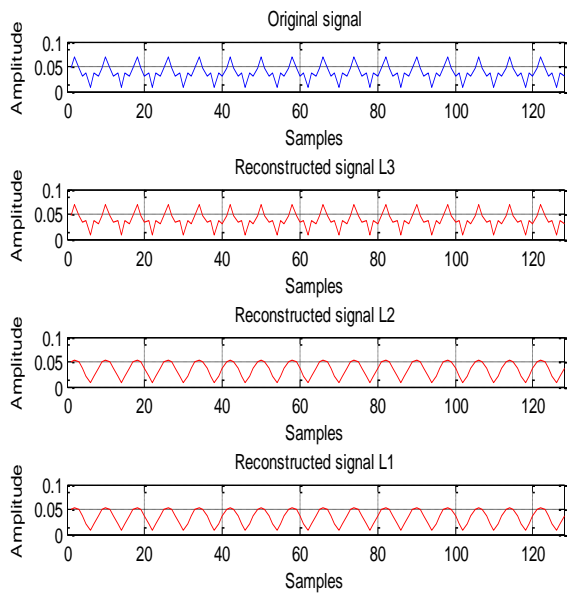

Fig. 5. Original and reconstructed signal in time domain using $l_3$, $l_2$ and $l_1$ minimization norms

From these figures we have the same concludion as we had in example 1 that norm $l_3$ gives the best reconstruction.

Beside this, for both noisy signals from example 1 and 2, except changing minimization norms, we also changed treshold. Treshold $\alpha \max\{V(k)\}$, $\alpha \operatorname{median}\{V(k)\}$ and $\alpha \operatorname{mean}\{V(k)\}$ is tested but we got similar results when parameter α is α=0.89.

## V. CONCLUSION

The non-iterative CS algorithm for signal reconstruction is considered. This algorithm is tested for signal corruped by Cauchy and Cubic Gaussian noises. In accordance with the nature of considered noises, we found that $l_3$ minimization norm provides accurate results. The achieved results are compared with the results obtained using $l_1$ and $l_2$ minimization norms. In this way, the presented theory and results proved that by using only minor modification, the same algorithm can run using various minimization norms.

ACKNOWLEDGEMENT

The authors are thankful to Professors and assistants within the Laboratory for Multimedia Signals and Systems, at the University of Montenegro, for providing the ideas, codes, literature and results developed for the project CS-ICT (funded by the Montenegrin Ministry of Science).